\documentclass[10pt,a4paper,onecolumn]{article}
\usepackage{marginnote}
\usepackage{graphicx}
\usepackage{xcolor}
\usepackage{authblk,etoolbox}
\usepackage{titlesec}
\usepackage{calc}
\usepackage{tikz}
\usepackage{hyperref}
\hypersetup{colorlinks,breaklinks,
            colorlinks=[rgb]{0.0, 0.5, 1.0}, 
            citecolor=[rgb]{0.0, 0.5, 1.0},
            urlcolor=[rgb]{0.0, 0.5, 1.0},
            linkcolor=[rgb]{0.0, 0.5, 1.0}}
\usepackage{caption}
\usepackage{amssymb,amsmath}
\usepackage{ifxetex,ifluatex}
\usepackage{fixltx2e} % provides \textsubscript
\usepackage{natbib}
\usepackage{bookmark}

\usepackage[top=3.5cm, bottom=3cm, right=1.5cm, left=1.0cm,
            headheight=2.2cm, reversemp, includemp, marginparwidth=4.5cm]{geometry}

\titleformat{\section}
  {\normalfont\sffamily\Large\bfseries}
  {}{0pt}{}
\titleformat{\subsection}
  {\normalfont\sffamily\large\bfseries}
  {}{0pt}{}
\titleformat{\subsubsection}
  {\normalfont\sffamily\bfseries}
  {}{0pt}{}
\titleformat*{\paragraph}
  {\sffamily\normalsize}

\usepackage{fancyhdr}
\makeatletter
\let\ps@plain\ps@fancy
\fancyheadoffset[L]{4.5cm}
\fancyfootoffset[L]{4.5cm}

\definecolor{linky}{rgb}{0.0, 0.5, 1.0}

\newcommand{\ExternalLink}{%
   \tikz[x=1.2ex, y=1.2ex, baseline=-0.05ex]{%
       \begin{scope}[x=1ex, y=1ex]
           \clip (-0.1,-0.1)
               --++ (-0, 1.2)
               --++ (0.6, 0)
               --++ (0, -0.6)
               --++ (0.6, 0)
               --++ (0, -1);
           \path[draw,
               line width = 0.5,
               rounded corners=0.5]
               (0,0) rectangle (1,1);
       \end{scope}
       \path[draw, line width = 0.5] (0.5, 0.5)
           -- (1, 1);
       \path[draw, line width = 0.5] (0.6, 1)
           -- (1, 1) -- (1, 0.6);
       }
   }

\patchcmd{\@maketitle}{center}{flushleft}{}{}
\patchcmd{\@maketitle}{center}{flushleft}{}{}
\patchcmd{\@maketitle}{\LARGE}{\LARGE\sffamily}{}{}
\def\maketitle{{%
  
  \AB@maketitle}}
\makeatletter
\renewcommand\AB@affilsepx{ \protect\Affilfont}
\renewcommand\AB@affilnote[1]{{\bfseries #1}\hspace{3pt}}
\makeatother

\renewcommand\Affilfont{\sffamily\small\mdseries}
\ifnum 0\ifxetex 1\fi\ifluatex 1\fi=0 % if pdftex
  \usepackage[T1]{fontenc}
  \usepackage[utf8]{inputenc}

\else % if luatex or xelatex
  \ifxetex
    \usepackage{mathspec}
  \else
    \usepackage{fontspec}
  \fi
  \defaultfontfeatures{Ligatures=TeX,Scale=MatchLowercase}

\fi
\IfFileExists{upquote.sty}{\usepackage{upquote}}{}
\IfFileExists{microtype.sty}{%
\usepackage{microtype}
\UseMicrotypeSet[protrusion]{basicmath} % disable protrusion for tt fonts
}{}

\usepackage{hyperref}
\hypersetup{unicode=true,
            pdftitle={AtomNeb: IDL Library and Python Package for Atomic Data of Ionized Nebulae},
            pdfborder={0 0 0},
            breaklinks=true}
\usepackage{graphicx,grffile}
\makeatletter
\def\maxwidth{\ifdim\Gin@nat@width>\linewidth\linewidth\else\Gin@nat@width\fi}
\def\maxheight{\ifdim\Gin@nat@height>\textheight\textheight\else\Gin@nat@height\fi}
\makeatother
\setkeys{Gin}{width=\maxwidth,height=\maxheight,keepaspectratio}
\IfFileExists{parskip.sty}{%
\usepackage{parskip}
}{% else
\setlength{\parindent}{0pt}
\setlength{\parskip}{6pt plus 2pt minus 1pt}
}
\ifx\paragraph\undefined\else
\let\oldparagraph\paragraph
\renewcommand{\paragraph}[1]{\oldparagraph{#1}\mbox{}}
\fi
\ifx\subparagraph\undefined\else
\let\oldsubparagraph\subparagraph
\renewcommand{\subparagraph}[1]{\oldsubparagraph{#1}\mbox{}}
\fi

\title{AtomNeb: IDL Library for Atomic Data of Ionized Nebulae}

        \author[1, 2, 3]{Ashkbiz Danehkar}
    
      \affil[1]{Research Centre in Astronomy, Astrophysics \& Astrophotonics, Macquarie University, Sydney, NSW 2109, Australia}
      \affil[2]{Harvard-Smithsonian Center for Astrophysics, 60 Garden Street, Cambridge, MA 02138, USA}
      \affil[3]{Department of Astronomy, University of Michigan, 1085 S. University Avenue, Ann Arbor, MI 48109, USA}
  \date{\vspace{-5ex}}

\begin{document}
\bookmark[page=1,level=0]{AtomNeb IDL Library}
\maketitle

\marginpar{
  %\hrule
  \sffamily\small

  {\bfseries DOI:} \href{https://doi.org/10.21105/joss.00898}{\color{linky}{10.21105/joss.00898}}

  \vspace{2mm}

  {\bfseries Software}
  \begin{itemize}
    \setlength\itemsep{0em}
    \item \href{https://github.com/openjournals/joss-reviews/issues/898}{\color{linky}{Review}} \ExternalLink
    \item \href{https://github.com/atomneb/AtomNeb-idl}{\color{linky}{Repository}} \ExternalLink
    \item \href{https://doi.org/10.5281/zenodo.2584420}{\color{linky}{Archive}} \ExternalLink
  \end{itemize}

  \vspace{2mm}

  {\bfseries Submitted:} 29 June 2018\\
  {\bfseries Published:} 06 March 2019

  \vspace{2mm}
  {\bfseries License}\\
  Authors of papers retain copyright and release the work under a Creative Commons Attribution 4.0 International License (\href{http://creativecommons.org/licenses/by/4.0/}{\color{linky}{CC-BY}}).
}

\vspace{8mm}
  
\hypertarget{summary}{%
\section{Summary}\label{summary}}

Ionized gaseous nebulae are interstellar clouds of hydrogen-rich
materials, which are photo-ionized by ultraviolet radiation from stars,
making them visible in multi-wavelength bands. Ionized nebulae can be
used as an astrophysical tool to trace the chemical composition of the
interstellar medium in our Galaxy and other galaxies, and to study
mixing processes in stellar evolution. Spectra emitted from ionized
nebulae generally contain collisionally excited and recombination lines.
Electron temperatures, electron densities, and ionic abundances can be
determined from \emph{collisionally excited lines} (CEL) by solving
statistical equilibrium equations using collision strengths
(\(\Omega_{ij}\)) and transition probabilities (\(A_{ij}\)) of ions.
Moreover, physical conditions and chemical abundances can be calculated
from \emph{recombination lines} (RL) using effective recombination
coefficients (\(\alpha_{\rm eff}\)) of ions. The atomic data,
i.e.~\(\Omega_{ij}\), \(A_{ij}\), and \(\alpha_{\rm eff}\), are used to
calculate line emissivities in nebular spectral analysis tools 
\citep[e.g.,][]{Howarth:1981,Shaw:1994,Shaw:1998,Luridiana:2015,Howarth:2016,Danehkar:2018a},
and photoionization codes \citep[e.g.,][]{Ferland:1998,Kallman:2001,Ercolano:2003,Ercolano:2005,Ercolano:2008}. Hence, the atomic data for collisional
excitation and recombination process are essential to determine physical
conditions and elemental abundances of ionized nebulae from
collisionally excited and recombination lines \citep[see, e.g.,][]{Danehkar:2013,Danehkar:2014,Danehkar:2014b,Danehkar:2016,Danehkar:2018b}.

\texttt{AtomNeb} is a database containing atomic data stored in the
Flexible Image Transport System (FITS) file format \citep{Wells:1981,Hanisch:2001,Pence:2010} produced for nebular spectral
analysis. FITS tables provide easy access to atomic data for spectral
analysis tools. Especially, \texttt{AtomNeb} includes the atomic data
for both the \emph{collisional excitation} and \emph{recombination}
process of ions usually observed in nebular astrophysics. The
\texttt{AtomNeb} interface library is equipped with several application
programming interface (API) functions written in the Interactive Data
Language (IDL) for reading the atomic data from the \texttt{AtomNeb}
FITS files. Furthermore, the \texttt{AtomNeb} IDL library can be
employed in the GNU Data Language (GDL) \citep[][]{Arabas:2010,Coulais:2010}, an open-source free compiler for IDL codes.

\begin{itemize}
\item
  The API functions for the \emph{CEL atomic data} developed in the IDL
  programming language were designed to easily read \emph{collision
  strengths} (\(\Omega_{ij}\)) and \emph{transition probabilities}
  (\(A_{ij}\)) of given ions, which can be used to derive electron
  temperatures, electron densities, and ionic abundances from measured
  fluxes of collisionally excited lines. The CEL data include energy
  levels (\(E_{j}\)), collision strengths (\(\Omega_{ij}\)), and
  transition probabilities (\(A_{ij}\)) from the CHIANTI database
  version 5.2 \citep{Landi:2006}, version 6.0 \citep{Dere:2009}, and
  version 7.0 \citep{Landi:2012}, which were compiled according to the
  atomic data used in the FORTRAN program \texttt{MOCASSIN}
  \citep{Ercolano:2003,Ercolano:2005,Ercolano:2008}.
  The CEL data also include a collection compiled based on the atomic
  data used in the Python package \texttt{pyNeb} for spectral analysis
  \citep{Luridiana:2015}.
\item
  The API functions for the \emph{RL atomic data} developed in IDL were
  designed to provide easy access to \emph{effective recombination
  coefficients} (\(\alpha_{\rm eff}\)) and \emph{branching ratios}
  (\(Br\)) of recombination lines of given ions. The RL data include
  effective recombination coefficients for C II \citep{Davey:2000}, N
  II \citep{Escalante:1990}, O II \citep{Storey:1994,Liu:1995}, and 
  Ne II \citep{Kisielius:1998}, which were
  compiled based on the atomic data in \texttt{MOCASSIN}. The RL data
  also include hydrogenic ions for Z=1 to 8 \citep{Storey:1995},
  effective recombination coefficients for H, He, C, N, O, and Ne ions
  \citep{Pequignot:1991}, effective recombination coefficients for He
  I \citep{Porter:2012,Porter:2013}, effective recombination
  coefficients for N II \citep{Fang:2011,Fang:2013}, and
  effective recombination coefficients for O II \citep{Storey:2017}.
\end{itemize}

The \texttt{AtomNeb} IDL/GDL package uses the FITS file related IDL
procedures from the IDL Astronomy User's library \citep{Landsman:1993,Landsman:1995} 
to read the atomic data from the \texttt{AtomNeb}
FITS files. The API functions of the \texttt{AtomNeb} IDL library,
together with the \texttt{proEQUIB} IDL library
\citep{Danehkar:2018a}, can be used to perform plasma diagnostics and
abundance analysis of nebular spectra emitted from ionized gaseous
nebulae.

\hypertarget{acknowledgements}{%
\section{Acknowledgements}\label{acknowledgements}}

A.D. acknowledges the receipt of a Macquarie University Research
Excellence Scholarship.

\bookmark[page=2,level=1]{References}

\newpage
\setcounter{page}{1}
\bookmark[page=5,level=0]{AtomNeb Python Package}

\fancyfoot[L]{\footnotesize{\sffamily Danehkar, (2020). AtomNeb Python Package, an addendum to AtomNeb: IDL Library for Atomic Data of Ionized Nebulae. \textit{Journal of Open Source Software}, 5(55), 2797. \href{https://doi.org/10.21105/joss.02797}{https://doi.org/10.21105/joss.02797}}}

\vspace{12pt}
\noindent {\LARGE\sffamily~\\~\\~\\\vspace{5pt}AtomNeb Python Package, an addendum to AtomNeb: IDL Library for Atomic Data of Ionized Nebulae}
\begin{flushleft}
\vspace{8pt}
{{\large{\sffamily{\bfseries Ashkbiz Danehkar}}$^{\text{1, 2, 3}}$}~\vspace{6pt}\\}
{\sffamily\small\mdseries \textbf{1} Department of Physics and Astronomy, Macquarie University, Sydney, NSW 2109, Australia} 
{\sffamily\small\mdseries \textbf{2} Harvard-Smithsonian Center for Astrophysics, 60 Garden Street, Cambridge, MA 02138, USA} 
{\sffamily\small\mdseries \textbf{3} Department of Astronomy, University of Michigan, 1085 S. University Avenue, Ann Arbor, MI 48109, USA}
\end{flushleft}

\maketitle

\marginpar{
 \vspace{0mm}
  %\hrule
  \sffamily\small

  {\bfseries DOI:} \href{https://doi.org/10.21105/joss.02797}{\color{linky}{10.21105/joss.02797}}

  \vspace{2mm}

  {\bfseries Software}
  \begin{itemize}
    \setlength\itemsep{0em}
    \item \href{https://github.com/openjournals/joss-reviews/issues/2797}{\color{linky}{Review}} \ExternalLink
    \item \href{https://github.com/atomneb/AtomNeb-py}{\color{linky}{Repository}} \ExternalLink
    \item \href{https://doi.org/10.5281/zenodo.4287566}{\color{linky}{Archive}} \ExternalLink
  \end{itemize}

  \vspace{2mm}

  {\bfseries Submitted:} 20 October 2020\\
  {\bfseries Published:} 24 November 2020

  \vspace{2mm}
  {\bfseries License}\\
  Authors of papers retain copyright and release the work under a Creative Commons Attribution 4.0 International License (\href{http://creativecommons.org/licenses/by/4.0/}{\color{linky}{CC BY 4.0}}).
}

\vspace{8mm}
  
\hypertarget{summary}{%
\section{Addendum}\label{addendum}}

\texttt{AtomNeb} is a Python open-source package containing atomic data
for gaseous nebulae stored in the Flexible Image Transport System (FITS)
file format \citep{ref-Wells:1981,ref-Hanisch:2001,ref-Pence:2010}. These FITS
files offer easy access to the atomic data required for emissivity
calculations in the collisional excitation and recombination processes
usually occurred in ionized gases of planetary nebulae and H II regions.
This package has several application programming interface (API)
functions developed in Python for retrieving the energy levels,
collision strengths, transition probabilities, and recombination
coefficients from its FITS files. The previous library \texttt{AtomNeb}
\citep{ref-Danehkar:2019} coupled to the library \texttt{proEQUIB}
\citep{ref-Danehkar:2018} needs the Interactive Data Language (IDL)
compiler, so this package offers an identical package for the high-level
programming language Python that can be used by those astrophysicists,
who intend to analyze nebular emission lines by developing codes in
Python. The \texttt{AtomNeb} Python functions can be used by the Python
package \texttt{pyEQUIB} \citep{ref-Danehkar:2020} to analyze emission-line
spectra.

\texttt{AtomNeb} uses the FITS handling routines of the Python package
\texttt{Astropy} \citep{ref-Astropy:2013,ref-Astropy:2018} to retrieve the
atomic data from its FITS files. It also requires the Python packages
\texttt{NumPy} \citep{ref-Walt:2011,ref-Harris:2020} and \texttt{pandas}
\citep{ref-McKinney:2010,ref-McKinney:2011,ref-McKinney:2017}. This package is
released under the GNU General Public License, and its source code is
publicly available on its GitHub repository. Its latest version can be
installed directly from its repository on the GitHub, and the stable
version from the Python Package Index (PyPi) via
\texttt{pip\ install\ atomneb} or alternatively from the Conda Python
package manager via \texttt{conda\ install\ -c\ conda-forge\ atomneb}.
The online documentation, tutorials and examples are provided on the
GitHub platform (\textsf{\url{https://github.com/atomneb/AtomNeb-py}}) and the Read the
Docs documentation host (\textsf{\url{https://atomneb-py.readthedocs.io/}}).

\hypertarget{acknowledgements}{%
\section{Acknowledgements}\label{acknowledgements}}

AD acknowledges the support of Research Excellence Scholarship from
Macquarie University.

%\hypertarget{references}{%
%\section*{References}\label{references}}


\begin{thebibliography}{38}

\bibitem[{Arabas} {et al.}(2010)]{Arabas:2010}
Arabas, S., M. Schellens, A. Coulais, J. Gales, and P. Messmer. 2010.
{``{GNU Data Language (GDL) - a free and open-source implementation of
IDL}.''} In \emph{EGU General Assembly Conference}, 12:924. Geophysical
Research Abstracts.

\bibitem[{Coulais} {et al.}(2010)]{Coulais:2010}
Coulais, A., M. Schellens, J. Gales, S. Arabas, M. Boquien, P. Chanial,
P. Messmer, et al. 2010. {``{Status of GDL - GNU Data Language}.''} In
\emph{Astronomical Data Analysis Software and Systems XIX}, edited by Y.
Mizumoto, K.-I. Morita, and M. Ohishi, 434:187. Astronomical Society of
the Pacific Conference Series. \url{https://arxiv.org/abs/1101.0679}.

\bibitem[{Danehkar}(2014)]{Danehkar:2014b}
Danehkar, A. 2014. {``{Evolution of Planetary Nebulae with WR-type
Central Stars}.''} PhD thesis, Macquarie University, Australia.
\url{https://doi.org/10.5281/zenodo.47794}.

\bibitem[{Danehkar}(2018a)]{Danehkar:2018a}
Danehkar, A. 2018a. {``{proEQUIB: IDL Library for Plasma Diagnostics and
Abundance Analysis}.''} \emph{The Journal of Open Source Software} 3 (32):
899. \url{https://doi.org/10.21105/joss.00899}.

\bibitem[{Danehkar}(2018b)]{Danehkar:2018b}
Danehkar, A. 2018b. {``{Bi-Abundance Ionisation Structure of the
Wolf-Rayet Planetary Nebula PB 8}.''} \emph{Publications of the
Astronomical Society of Australia} 35: e005.
\url{https://doi.org/10.1017/pasa.2018.1}.

\bibitem[{Danehkar} {et al.}(2013)]{Danehkar:2013}
Danehkar, A., Q. A. Parker, and B. Ercolano. 2013. {``{Observations and
three-dimensional ionization structure of the planetary nebula SuWt
2}.''} \emph{Monthly Notices of the Royal Astronomical Society} 434:
1513--30. \url{https://doi.org/10.1093/mnras/stt1116}.

\bibitem[{Danehkar} {et al.}(2016)]{Danehkar:2016}
Danehkar, A., Q. A. Parker, and W. Steffen. 2016. {``{Fast,
Low-ionization Emission Regions of the Planetary Nebula M2-42}.''}
\emph{The Astronomical Journal} 151: 38.
\url{https://doi.org/10.3847/0004-6256/151/2/38}.

\bibitem[{Danehkar} {et al.}(2014)]{Danehkar:2014}
Danehkar, A., H. Todt, B. Ercolano, and A. Y. Kniazev. 2014.
{``{Observations and three-dimensional photoionization modelling of the
Wolf-Rayet planetary nebula Abell 48}.''} \emph{Monthly Notices of the
Royal Astronomical Society} 439: 3605--15.
\url{https://doi.org/10.1093/mnras/stu203}.

\bibitem[{Davey} {et al.}(2000)]{Davey:2000}
Davey, A. R., P. J. Storey, and R. Kisielius. 2000. {``{Recombination
coefficients for C II lines}.''} \emph{Astronomy and Astrophysicss} 142:
85--94. \url{https://doi.org/10.1051/aas:2000139}.

\bibitem[{Dere} {et al.}(2009)]{Dere:2009}
Dere, K. P., E. Landi, P. R. Young, G. Del Zanna, M. Landini, and H. E.
Mason. 2009. {``{CHIANTI - an atomic database for emission lines. IX.
Ionization rates, recombination rates, ionization equilibria for the
elements hydrogen through zinc and updated atomic data}.''}
\emph{Astronomy and Astrophysics} 498: 915--29.
\url{https://doi.org/10.1051/0004-6361/200911712}.

\bibitem[{Ercolano} {et al.}(2005)]{Ercolano:2005}
Ercolano, B., M. J. Barlow, and P. J. Storey. 2005. {``{The dusty
MOCASSIN: fully self-consistent 3D photoionization and dust radiative
transfer models}.''} \emph{Monthly Notices of the Royal Astronomical
Society} 362: 1038--46.
\url{https://doi.org/10.1111/j.1365-2966.2005.09381.x}.

\bibitem[{Ercolano} {et al.}(2003)]{Ercolano:2003}
Ercolano, B., M. J. Barlow, P. J. Storey, and X.-W. Liu. 2003.
{``{MOCASSIN: a fully three-dimensional Monte Carlo photoionization
code}.''} \emph{Monthly Notices of the Royal Astronomical Society} 340:
1136--52. \url{https://doi.org/10.1046/j.1365-8711.2003.06371.x}.

\bibitem[{Ercolano} {et al.}(2008)]{Ercolano:2008}
Ercolano, B., P. R. Young, J. J. Drake, and J. C. Raymond. 2008.
{``{X-Ray Enabled MOCASSIN: A Three-dimensional Code for Photoionized
Media}.''} \emph{The Astrophysical Journal Supplement Series} 175:
534--42. \url{https://doi.org/10.1086/524378}.

\bibitem[{Escalante \& Victor}(1990)]{Escalante:1990}
Escalante, V., and G. A. Victor. 1990. {``{Effective recombination
coefficients of neutral carbon and singly ionized nitrogen}.''}
\emph{The Astrophysical Journal Supplement Series} 73: 513--53.
\url{https://doi.org/10.1086/191479}.

\bibitem[{Fang} {et al.}(2011)]{Fang:2011}
Fang, X., P. J. Storey, and X.-W. Liu. 2011. {``{New effective
recombination coefficients for nebular N ii lines}.''} \emph{Astronomy
and Astrophysics} 530: A18.
\url{https://doi.org/10.1051/0004-6361/201116511}.

\bibitem[{Fang} {et al.}(2013)]{Fang:2013}
Fang, X., P. J. Storey, and X.-W. Liu. 2013. {``{New effective recombination coefficients for
nebular N II lines (Corrigendum)}.''} \emph{Astronomy and Astrophysics}
550: C2. \url{https://doi.org/10.1051/0004-6361/201116511e}.

\bibitem[{Ferland} {et al.}(1998)]{Ferland:1998}
Ferland, G. J., K. T. Korista, D. A. Verner, J. W. Ferguson, J. B.
Kingdon, and E. M. Verner. 1998. {``{CLOUDY 90: Numerical Simulation of
Plasmas and Their Spectra}.''} \emph{Publications of the Astronomical
Society of the Pacific} 110: 761--78.
\url{https://doi.org/10.1086/316190}.

\bibitem[{Hanisch} {et al.}(2001)]{Hanisch:2001}
Hanisch, R. J., A. Farris, E. W. Greisen, W. D. Pence, B. M.
Schlesinger, P. J. Teuben, R. W. Thompson, and A. Warnock III. 2001.
{``{Definition of the Flexible Image Transport System (FITS)}.''}
\emph{Astronomy and Astrophysics} 376: 359--80.
\url{https://doi.org/10.1051/0004-6361:20010923}.

\bibitem[{Howarth \& Adams}(1981)]{Howarth:1981}
Howarth, I. D., and S. Adams. 1981. {``{Program EQUIB}.''} University
College London.

\bibitem[{Howarth} {et al.}(2016)]{Howarth:2016}
Howarth, I. D., S. Adams, R. E. S. Clegg, D. P. Ruffle, X.-W. Liu, C. J.
Pritchet, and B. Ercolano. 2016. {``{EQUIB: Atomic level populations and
line emissivities calculator}.''} Astrophysics Source Code Library,
ascl:1603.005.

\bibitem[{Kallman \& Bautista}(2001)]{Kallman:2001}
Kallman, T., and M. Bautista. 2001. {``{Photoionization and High-Density
Gas}.''} \emph{The Astrophysical Journal Supplement Series} 133:
221--53. \url{https://doi.org/10.1086/319184}.

\bibitem[{Kisielius} {et al.}(1998)]{Kisielius:1998}
Kisielius, R., P. J. Storey, A. R. Davey, and L. T. Neale. 1998.
{``{Recombination coefficients for Ne Ii lines at nebular temperatures
and densities}.''} \emph{Astronomy and Astrophysicss} 133: 257--69.
\url{https://doi.org/10.1051/aas:1998319}.

\bibitem[{Landi} {et al.}(2012)]{Landi:2012}
Landi, E., G. Del Zanna, P. R. Young, K. P. Dere, and H. E. Mason. 2012.
{``{CHIANTI--An Atomic Database for Emission Lines. XII. Version 7
of the Database}.''} \emph{The Astrophysical Journal} 744: 99.
\url{https://doi.org/10.1088/0004-637X/744/2/99}.

\bibitem[{Landi} {et al.}(2006)]{Landi:2006}
Landi, E., G. Del Zanna, P. R. Young, K. P. Dere, H. E. Mason, and M.
Landini. 2006. {``{CHIANTI-An Atomic Database for Emission Lines. VII.
New Data for X-Rays and Other Improvements}.''} \emph{The Astrophysical
Journal Supplement Series} 162: 261--80.
\url{https://doi.org/10.1086/498148}.

\bibitem[{Landsman}(1993)]{Landsman:1993}
Landsman, W. B. 1993. {``{The IDL Astronomy User's Library}.''} In
\emph{Astronomical Data Analysis Software and Systems II}, edited by R.
J. Hanisch, R. J. V. Brissenden, and J. Barnes, 52:246. Astronomical
Society of the Pacific Conference Series.

\bibitem[{Landsman}(1995)]{Landsman:1995}
Landsman, W. B. 1995. {``{The IDL Astronomy User's Library}.''} In
\emph{Astronomical Data Analysis Software and Systems IV}, edited by R.
A. Shaw, H. E. Payne, and J. J. E. Hayes, 77:437. Astronomical Society
of the Pacific Conference Series.

\bibitem[{Liu} {et al.}(1995)]{Liu:1995}
Liu, X.-W., P. J. Storey, M. J. Barlow, and R. E. S. Clegg. 1995.
{``{The rich O II recombination spectrum of the planetary nebula NGC
7009: new observations and atomic data}.''} \emph{Monthly Notices of the
Royal Astronomical Society} 272: 369--88.
\url{https://doi.org/10.1093/mnras/272.2.369}.

\bibitem[{Luridiana} {et al.}(2015)]{Luridiana:2015}
Luridiana, V., C. Morisset, and R. A. Shaw. 2015. {``{PyNeb: a new tool
for analyzing emission lines. I. Code description and validation of
results}.''} \emph{Astronomy and Astrophysics} 573: A42.
\url{https://doi.org/10.1051/0004-6361/201323152}.

\bibitem[{Pence} {et al.}(2010)]{Pence:2010}
Pence, W. D., L. Chiappetti, C. G. Page, R. A. Shaw, and E. Stobie.
2010. {``{Definition of the Flexible Image Transport System (FITS),
version 3.0}.''} \emph{Astronomy and Astrophysics} 524: A42.
\url{https://doi.org/10.1051/0004-6361/201015362}.

\bibitem[{Pequignot} {et al.}(1991)]{Pequignot:1991}
Pequignot, D., P. Petitjean, and C. Boisson. 1991. {``{Total and
effective radiative recombination coefficients}.''} \emph{Astronomy and
Astrophysics} 251: 680--88.

\bibitem[{Porter} {et al.}(2012)]{Porter:2012}
Porter, R. L., G. J. Ferland, P. J. Storey, and M. J. Detisch. 2012.
{``{Improved He I emissivities in the case B approximation}.''}
\emph{Monthly Notices of the Royal Astronomical Society} 425: L28--31.
\url{https://doi.org/10.1111/j.1745-3933.2012.01300.x}.

\bibitem[{Porter} {et al.}(2013)]{Porter:2013}
Porter, R. L., G. J. Ferland, P. J. Storey, and M. J. Detisch. 2013. 
{``{Erratum: {`Improved He I Emissivities in the Case B
Approximation'}}.''} \emph{Monthly Notices of the Royal Astronomical
Society} 433: L89--90. \url{https://doi.org/10.1093/mnrasl/slt049}.

\bibitem[{Shaw \& Dufour}(1994)]{Shaw:1994}
Shaw, R. A., and R. J. Dufour. 1994. {``{The FIVEL Nebular Modelling
Package in STSDAS}.''} In \emph{Astronomical Data Analysis Software and
Systems III}, edited by D. R. Crabtree, R. J. Hanisch, and J. Barnes,
61:327. Astronomical Society of the Pacific Conference Series.

\bibitem[{Shaw} {et al.}(1998)]{Shaw:1998}
Shaw, R. A., M. D. de La Pena, R. M. Katsanis, and R. E. Williams. 1998.
{``{Analysis Tools for Nebular Emission Lines}.''} In \emph{Astronomical
Data Analysis Software and Systems VII}, edited by R. Albrecht, R. N.
Hook, and H. A. Bushouse, 145:192. Astronomical Society of the Pacific
Conference Series.

\bibitem[{Storey}(1994)]{Storey:1994}
Storey, P. J. 1994. {``{Recombination coefficients for O II lines at
nebular temperatures and densities}.''} \emph{Astronomy and
Astrophysics} 282: 999--1013.

\bibitem[{Storey \& Hummer}(1995)]{Storey:1995}
Storey, P. J., and D. G. Hummer. 1995. {``{Recombination line
intensities for hydrogenic ions-IV. Total recombination coefficients and
machine-readable tables for Z=1 to 8}.''} \emph{Monthly Notices of the
Royal Astronomical Society} 272: 41--48.
\url{https://doi.org/10.1093/mnras/272.1.41}.

\bibitem[{Storey} {et al.}(2017)]{Storey:2017}
Storey, P. J., T. Sochi, and R. Bastin. 2017. {``{Recombination
coefficients for O II lines in nebular conditions}.''} \emph{Monthly
Notices of the Royal Astronomical Society} 470: 379--89.
\url{https://doi.org/10.1093/mnras/stx1189}.

\bibitem[{Wells} {et al.}(1981)]{Wells:1981}
Wells, D. C., E. W. Greisen, and R. H. Harten. 1981. {``{FITS - a
Flexible Image Transport System}.''} \emph{Astronomy and Astrophysics
Supplement} 44: 363.


\end{thebibliography}

\addcontentsline{toc}{section}{References}

\begin{thebibliography}{13}

\bibitem[{Astropy Collaboration} {et al.}(2018)]{ref-Astropy:2018} 
Astropy Collaboration, A. M. Price-Whelan, B. M. Sipőcz, H. M. Günther,
P. L. Lim, S. M. Crawford, S. Conseil, et al. 2018. {{The Astropy
Project: Building an Open-science Project and Status of the v2.0 Core
Package}.} \emph{The Astronomical Journal} 156 (3): 123.
\url{https://doi.org/10.3847/1538-3881/aabc4f}.

\bibitem[{Astropy Collaboration} {et al.}(2013)]{ref-Astropy:2013} 
Astropy Collaboration, T. P. Robitaille, E. J. Tollerud, P. Greenfield,
M. Droettboom, Erik Bray, Tom Aldcroft, et al. 2013. {{Astropy: A
community Python package for astronomy}.} \emph{Astronomy and
Astrophysics} 558: A33.
\url{https://doi.org/10.1051/0004-6361/201322068}.

\bibitem[{Danehkar}(2018)]{ref-Danehkar:2018}
Danehkar, A. 2018. {{proEQUIB: IDL Library for Plasma Diagnostics and
Abundance Analysis}.} \emph{The Journal of Open Source Software} 3 (32):
899. \url{https://doi.org/10.21105/joss.00899}.

\bibitem[{Danehkar}(2019)]{ref-Danehkar:2019}
Danehkar, A. 2019. {{AtomNeb: IDL Library for Atomic Data of Ionized
Nebulae}.} \emph{The Journal of Open Source Software} 4 (35): 898.
\url{https://doi.org/10.21105/joss.00898}.

\bibitem[{Danehkar}(2020)]{ref-Danehkar:2020}
Danehkar, A. 2020. {{pyEQUIB Python Package, an addendum to proEQUIB:
IDL Library for Plasma Diagnostics and Abundance Analysis}.} \emph{The
Journal of Open Source Software} 5 (55):
2798. \url{https://doi.org/10.21105/joss.02798}.

\bibitem[{Hanisch} {et al.}(2001)]{ref-Hanisch:2001}
Hanisch, R. J., A. Farris, E. W. Greisen, W. D. Pence, B. M.
Schlesinger, P. J. Teuben, R. W. Thompson, and A. Warnock III. 2001.
{{Definition of the Flexible Image Transport System (FITS)}.}
\emph{Astronomy and Astrophysics} 376: 359--80.
\url{https://doi.org/10.1051/0004-6361:20010923}.

\bibitem[{Harris} {et al.}(2020)]{ref-Harris:2020}
Harris, Charles R., K. Jarrod Millman, S. J. van der Walt, R. Gommers,
Pauli Virtanen, David Cournapeau, Eric Wieser, et al. 2020. {{Array
Programming with NumPy}.} \emph{Nature} 585: 357.
\url{https://doi.org/10.1038/s41586-020-2649-2}.

\bibitem[{McKinney}(2010)]{ref-McKinney:2010}
McKinney, W. 2010. {{Data Structures for Statistical Computing in
Python}.} In, 445:51--56. Proceedings of the 9th Python in Science
Conference (SciPy). \url{https://doi.org/10.25080/Majora-92bf1922-00a}.

\bibitem[{McKinney}(2011)]{ref-McKinney:2011}
McKinney, W. 2011. {{pandas: a Foundational Python Library for Data
Analysis and Statistics}.} In. Vol. 14. Python for High Performance
and Scientific Computing 9.

\bibitem[{McKinney}(2017)]{ref-McKinney:2017}
McKinney, W. 2017. \emph{{Python for Data Analysis: Data Wrangling with
Pandas, NumPy, and IPython, 2nd Edition}}. O'Reilly Media.

\bibitem[{Pence} {et al.}(2010)]{ref-Pence:2010}
Pence, W. D., L. Chiappetti, C. G. Page, R. A. Shaw, and E. Stobie.
2010. {{Definition of the Flexible Image Transport System (FITS),
version 3.0}.} \emph{Astronomy and Astrophysics} 524: A42.
\url{https://doi.org/10.1051/0004-6361/201015362}.

\bibitem[{van der Walt} {et al.}(2011)]{ref-Walt:2011}
van der Walt, S., S. C. Colbert, and G. Varoquaux. 2011. {{The NumPy
Array: A Structure for Efficient Numerical Computation}.}
\emph{Computing in Science and Engineering} 13 (2): 22--30.
\url{https://doi.org/10.1109/MCSE.2011.37}.

\bibitem[{Wells} {et al.}(1981)]{ref-Wells:1981}
Wells, D. C., E. W. Greisen, and R. H. Harten. 1981. {{FITS - a
Flexible Image Transport System}.} \emph{Astronomy and Astrophysics
Supplement} 44: 363.

\end{thebibliography}
\end{document}